\documentclass[aps,prl,twocolumn]{revtex4-2}

\usepackage{graphicx}% Include figure files
\usepackage{dcolumn}% Align table columns on decimal point
\usepackage{siunitx}
\usepackage{bm}% bold math
\usepackage{hyperref}% add hypertext capabilities
\hypersetup{colorlinks=true,linkcolor=blue,citecolor=blue,urlcolor=blue}
\usepackage[colorinlistoftodos]{todonotes}

\usepackage{lipsum}

\usepackage{lipsum} % nur für Beispieltext

\usepackage{setspace}

\begin{document}

\onecolumngrid

% Titel
\begin{center}
{\large\bfseries Unveiling Excitonic Insulator Signatures in Ta$_\mathrm{2}$NiSe$_\mathrm{5}$\\[2pt] 
 through Structural and Orbital Probes}\par
  \vspace{0.2em}
\end{center}

\begin{center}
\begin{minipage}{0.77\textwidth}
\centering
Nour Maraytta,${}^{1,*}$ Peter Nagel,${}^{1,2}$ Fatemeh Ghorbani,${}^{1}$ Amir Ghiami,${}^{1,2}$ Santanu Pakhira,${}^{1}$ Mai Ye,${}^{1}$ Björn Wehinger,${}^{3}$ Federico Abbruciati,${}^{3}$ Gaston Garbarino,${}^{3}$ Matthieu Le Tacon,${}^{1}$ Stefan Schuppler,${}^{1,2}$ Amir-Abbas Haghighirad,${}^{1}$ and Michael Merz${}^{1,2,*}$
\end{minipage}
\end{center}

\vspace{-1.5em}

\begin{center}
\begin{minipage}{1.0\textwidth}
\centering
{\small
\textit{ ${}^1$Institute for Quantum Materials and Technologies,\\ Karlsruhe Institute of Technology, Kaiserstr.\@ 12,\@ 76131 Karlsruhe, Germany\\
${}^2$Karlsruhe Nano Micro Facility (KNMFi), Karlsruhe Institute of Technology, Kaiserstr.\@ 12,\@ 76131 Karlsruhe, Germany\\  
${}^3$ESRF, The European Synchrotron, 71, avenue des Martyrs, CS 40220 F-38043 Grenoble Cedex 9, France}\\[1.25mm]
\textit{${}^*$Corresponding authors: nour.maraytta@kit.edu, michael.merz@kit.edu}\\[0.75mm]
(Dated: \today) }
\end{minipage}
\end{center}

\vspace{-1mm}

\begin{center}
\begin{minipage}{0.78\textwidth}
\setstretch{0.9}
\noindent
{\small
The high-temperature phase of Ta$_\mathrm{2}$NiSe$_\mathrm{5}$,\@ a near-zero-gap semiconductor ($E_G$ = 0), is a promising candidate for an excitonic insulator. Given the dome-like evolution expected for an excitonic insulator around $E_G$,\@ we investigated Ta$_\mathrm{2}$NiSe$_\mathrm{5}$,\@ the more semi-metallic Ta$_\mathrm{2}$(Ni,Co)Se$_\mathrm{5}$,\@ and semiconducting Ta$_\mathrm{2}$NiS$_\mathrm{5}$ using high-resolution single-crystal x-ray diffraction and near-edge x-ray absorption fine structure (NEXAFS). Our findings reveal a second-order structural phase transition from orthorhombic (space group:\@ $Cmcm$) to monoclinic (space group:\@ $C2/c$) in Ta$_\mathrm{2}$NiSe$_\mathrm{5}$ and Ta$_\mathrm{2}$(Ni,Co)Se$_\mathrm{5}$,\@ but no transition in Ta$_\mathrm{2}$NiS$_\mathrm{5}$ down to 2 K.\@ This transition breaks two mirror symmetries, enabling and enhancing the hybridization of Ta, Ni, and Se atoms, shortening bond lengths, and strengthening orbital interactions. NEXAFS data confirm stronger hybridization, significant changes in excitonic binding energies, and a key alteration in orbital character, suggesting an excitonic insulating state in Ta$_\mathrm{2}$NiSe$_\mathrm{5}$ and emphasizing the crucial electronic role of orbitals in the formation of the excitonic insulator state.
}

\end{minipage}
\end{center}

%\vspace{1.5em}
\vspace{2.1em}

\thispagestyle{empty}
\twocolumngrid
\setstretch{0.95}

\section{Introduction}

Quantum materials harbor a plethora of very intriguing exotic many-body phases among which one of the most fascinating phenomena 
is the emergence of an excitonic insulator (EI).\@ The idea of an EI was theoretically already proposed about half a century ago. It is well known that poorly screened Coulomb interactions between conduction-band electrons and valence-band holes can lead to the formation of electron-hole pairs called excitons \cite{Mott_6_1961, Keldysh_6_1965}. Especially, if these electron-hole pairs are formed in the presence of an almost vanishing band gap such charge neutral quasi-particles have the potential to condense in an unconventional insulating ground state, the EI \cite{Jerome_158_1967,Kohn_19_1967}.\@ Consequently, the EI was predicted to be realized in narrow band gap semiconductors (semimetals) with a small positive (negative) band gap. 
Theoretically, it has been established that the BCS theory can be used to describe the EI phase in semimetals, while for semiconductors it can be better described in terms of Bose-Einstein condensation (BEC) \cite{Bronold_74_2006, Ihle_78_2008, Phan_81_2010}.

Candidate materials for possible realizations of an EI include TmSe$_{1-x}$Te$_x$, where it has been claimed, based on transport properties, that the pressure induced transition in this compound points to an EI state \cite{Neuenschwander_41_1990, Bucher_67_1991, Wachter_118_2001}. An alternative candidate is the semi-metallic compound 1\textit{T}-TiSe$_\mathrm{2}$, where a chiral charge density wave (CDW) was observed below $T_\mathrm{CDW}$ $\approx$ 200 K \cite{Kim2024} and interpreted to be of the EI type \cite{Cercellier_99_2007, Wilson_22_1977, Maschek_94_2016}. However, in contrast to the pure EI scenario where lattice distortions play no or merely a secondary role, the transitions in both compounds are strongly influenced by lattice 
degrees of freedom, thereby, obscuring a definitive assessment and leaving room for questioning the existence of an EI phase in these materials \cite{Bucher_2008, Craven_25_1978, Lu_8_2017}. 
 
The transition-metal chalcogenide Ta$_\mathrm{2}$NiSe$_\mathrm{5}$, which was first reported by Sunshine et al. \cite{sunshine_24_1985} and Di Salvo et al. \cite{SALVO_116_1986},\@ is considered a further promising candidate for the realization of the EI state.\@ This compound has a layered structure for which \textit{ac} planes composed of one Ni-Se and two Ta-Se chains along the crystallographic \textit{a} axis are stacked by van der Waals interactions along the \textit{b} direction, 
as illustrated in Fig.\@ \ref{fig: hybridization model}.\@ Within the chains, Ni (Ta) is tetrahedrally (octahedrally) coordinated with Se ions. The compound undergoes a second order structural phase transition from orthorhombic, space group (SG) \textit{Cmcm}, to monoclinic, SG \textit{C2/c},\@ at $T_\mathrm{C}$ $\approx$ 328 K \cite{sunshine_24_1985, SALVO_116_1986}.\@ A semiconductor- or semimetal-to-insulator (SI) transition is observed at $T_\mathrm{C}$ and has been assumed to be driven by an EI origin \cite{ Lu_8_2017, Lee_99_2019,  Fukutani_123_2019, Kim_12_2021}. 

First indications for an EI ground state were found in Ta$_\mathrm{2}$NiSe$_\mathrm{5}$ using optical conductivity measurements where an opening of a gap of $\approx$ 0.16 eV, comparable to the expected excitonic binding energy consistent with the enhanced activation energy observed in the transport data and the rapid increase in the resistivity below $T_\mathrm{C}$ \cite{ Lu_8_2017}.\@ Moreover,
angle resolved photoemission spectroscopy (ARPES) shows an M-shaped flattening of the valence band top below  $T_\mathrm{C}$, which was interpreted as the formation of an additional gap induced by the EI transition \cite{Wakisaka_103_2009}.\@ This result was also supported by band structure calculations where a very small direct gap at the $\Gamma$ point of the Brillouin zone is reported above $T_\mathrm{C}$.\@ No superlattice reflections are observed below $T_\mathrm{C}$, which excludes a CDW as the origin for the structural transition in Ta$_\mathrm{2}$NiSe$_\mathrm{5}$ \cite{Kaneko_87_2013}.\@ Furthermore, the leading role of electronic correlations in driving the EI transition is supported by several theoretical studies \cite{Kaneko_87_2013, Sugimoto_93_2016, Sugimoto_120_2018, Yamada_85_2016} and by the observation of critical charge fluctuations in Raman spectra \cite{Kim_12_2021, Katsumi_130_2023}. Nevertheless, there is an ongoing debate about whether the transition is primarily of excitonic origin \cite{Lu_8_2017, Lee_99_2019, Fukutani_123_2019, Kim_12_2021, Wakisaka_103_2009, Kaneko_87_2013, Sugimoto_93_2016, Sugimoto_120_2018, Yamada_85_2016, Katsumi_130_2023, Mazza_124_2020} or driven by lattice distortion and electron-phonon coupling \cite{Watson_2_2020, Baldini_120_2023, Chen_5_2023}.

Previous investigations on Ta$_\mathrm{2}$Ni(Se,S)$_\mathrm{5}$, where Se is replaced by S,\@ revealed that increasing S substitution weakens the hybridization between Ni $3d$ and Se $4p$ (S $3p$) orbitals, consequently enhancing the band gap energy, $E_\mathrm{G}$. Specifically, the critical temperature, $T_C$,\@ decreases with increasing $x$ and disappears around $x \approx 0.7$,\@ while $E_\mathrm{G}$ exhibits a linear increase as a function of $x$ \cite{Lu_8_2017, Sano_30_2020}. For the quasi-1D end member of the series, Ta$_2$NiS$_5$,\@ no phase transition was observed in the temperature-dependent resistivity down to 2 K.\@ Furthermore, both transport and optical measurements indicate that $E_\mathrm{G}$ exceeds the binding energy of the exciton, $E_\mathrm{B}$,\@ suggesting the absence of an excitonic transition in Ta$_2$NiS$_5$ \cite{Lu_8_2017, Larkin_2016}.\@ Additionally, ARPES revealed no evidence for a flattening at the top of the valence band, indicating that the ground state of an EI cannot be realized in this material  \cite{mu_6_2018}. In contrast, substitution of Co for Ni in Ta$_\mathrm{2}$(Ni,Co)Se$_\mathrm{5}$ increases the negative band gap and reduces the transition temperature, reinforcing the semimetallic character. As shown in Ref. \cite{Mitsuoka_89_2020},\@ the transition temperature significantly decreases to 250 K for a Co content around 10 \%, and disappears at around $x$ = 0.27 \cite{Hirose_92_2023}.

Hence, through appropriate S and Co substitution in Ta$_\mathrm{2}$NiSe$_\mathrm{5}$,\@ $T_\mathrm{C}$ is suppressed by directing the system towards either a semiconducting (positive band gap) or a more semimetallic state (negative band gap),\@ resulting in a substitution-dependent dome-shaped behavior centered at $E_\mathrm{G}$ = 0,\@ as anticipated for an excitonic insulator (EI) \cite{Lu_8_2017}.\@ Along these lines, the primary motivation for our present investigation lies in probing the spatial and electronic stucture of Ta$_\mathrm{2}$(Ni,Co)(Se,S)$_\mathrm{5}$ employing temperature-dependent high-resolution single-crystal x-ray diffraction (SC-XRD) as well as near-edge x-ray absorption fine structure (NEXAFS),\@ to scrutinize the existence of an EI state in Ta$_\mathrm{2}$NiSe$_\mathrm{5}$ and, if the EI state indeed exists, to identify the relevant Ta, Ni, and Se orbitals responsible for its formation.

We will demonstrate that not only Ta$_\mathrm{2}$NiSe$_\mathrm{5}$ but also Ta$_\mathrm{2}$Ni$_\mathrm{0.93}$Co$_\mathrm{0.07}$Se$_\mathrm{5}$ undergo a second-order structural phase transition from orthorhombic ($Cmcm$) to monoclinic ($C2/c$) symmetry. This transition is characterized by mirror symmetry breaking, which facilitates the hybridization of Ta, Ni, and Se atoms, leading to shortened bond lengths and enhanced orbital interactions. 
 In contrast, Ta$_\mathrm{2}$NiS$_\mathrm{5}$---in which numerous studies have been conducted, leading to various interpretations and controversies about the existence of a phase transition \cite{sunshine_24_1985, Lu_8_2017, SALVO_116_1986, Sano_30_2020, Ye_104_2021, Ye_110_2024, Windgätter_7_2021}---shows no evidence of symmetry breaking or changes in bond lengths. The structural findings are corroborated by NEXAFS data, which reveal changes in orbital symmetries and hybridizations, unequivocally indicating the formation of an EI state. This effect is most pronounced in Ta$_\mathrm{2}$NiSe$_\mathrm{5}$,\@ significantly reduced in Ta$_\mathrm{2}$Ni$_\mathrm{0.93}$Co$_\mathrm{0.07}$Se$_\mathrm{5}$,\@ and, in principle, absent in Ta$_\mathrm{2}$NiS$_\mathrm{5}$.

\section{Experimental Methods}

Single crystals of Ta$_\mathrm{2}$NiSe$_\mathrm{5}$, Ta$_\mathrm{2}$(Ni,Co)(Se,S)$_\mathrm{5}$, and Ta$_\mathrm{2}$NiS$_\mathrm{5}$ were synthesized using the chemical vapor transport method, as described in the Supplementary Material (SM) and shown in Fig.\@ S1 \cite{SM}.\@ The composition of the obtained samples was verified using energy-dispersive x-ray spectroscopy (EDX)---as illustrated in Fig.\@ S2 in the SM---and SC-XRD.\@ The Co content of the Ta$_\mathrm{2}$(Ni,Co)Se$_\mathrm{5}$ sample was determined to 7~\%.\@

Temperature-dependent SC-XRD data on Ta$_\mathrm{2}$NiSe$_\mathrm{5}$,\@ Ta$_\mathrm{2}$NiS$_\mathrm{5}$,\@ and Ta$_\mathrm{2}$Ni$_\mathrm{0.93}$Co$_\mathrm{0.07}$Se$_\mathrm{5}$ were collected between 360 and 80 K using our in-house high-flux, high-resolution, rotating anode Rigaku Synergy-DW (Mo/Ag) diffractometer with Mo $K_\mathrm{\alpha}$ radiation. The system is equipped with a background-less Hypix-Arc150$^{\circ}$ detector, which guarantees minimal reflection profile distortion and ensures uniform detection conditions for all reflections. All samples were measured to a resolution better than 0.5 \AA.\@ The samples exhibited no mosaic spread and no additional reflections from secondary phases, highlighting their high quality and allowing for excellent evaluation using the latest version of the CrysAlisPro software package \cite{CrysAlis}.\@ The crystal structures of the compounds were refined using JANA2006 \cite{Vaclav_229_2014},\@ including all averaged symmetry-independent reflections (I $>$ 2 $\sigma$) for the refinements in their respective space groups. For each measured temperature, the unit cell and space group were determined, atoms were localized within the unit cell using random phases and the structure was completed and solved using difference Fourier analysis. The structural refinements converged well for all investigated temperatures, exhibiting excellent reliability factors (see Tables S1 and S2 in the SM \cite{SM} for residuals $wR_\mathrm{2}$,\@ $R_\mathrm{1}$,\@ and goodness of fit, GOF,\@ values). For Ta$_2$NiS$_5$ we complemented the data by an SC-XRD experiment at ID15B of the European Synchrotron Radiation Facility (ESRF)  \cite{ID15B} down to a temperature of 2 K:\@ Samples were mounted with Apiezon N grease on a 300 $\mu$m thick diamond plate and were cooled with a Helium flow cryostat. Monochromatic x-rays with a wavelength of 0.41 {\AA} were used, with a spot size of $4 \times 4$ $\mu$m.\@ Data were collected through continuous $\phi$-rotation over a range of 70$^\circ$,\@ with shutterless readout at intervals of  0.5$^\circ$.\@ For the data acquisition an EIGER2 X CdTe 9M detector (Dectris, Switzerland) was employed.\@ The detector distance of 179.60 mm was determined utilizing CeO$_2$ powder and the geometry was calibrated with a natural vanadinite single crystal.

Before conducting our NEXAFS studies at the Institute for Quantum Materials and Technologies beamline WERA at the KIT light source KARA,\@ the orientation of the individual samples was precisely determined using a completely motorized Photonic Science Laue diffraction system (details of our Laue measurements can be found in Fig.\@ S7 in the SM \cite{SM}).\@ For Ta$_\mathrm{2}$NiSe$_\mathrm{5}$,\@ Ta$_\mathrm{2}$NiS$_\mathrm{5}$,\@ and Ta$_\mathrm{2}$Ni$_\mathrm{0.93}$Co$_\mathrm{0.07}$Se$_\mathrm{5}$,\@ temperature-dependent NEXAFS data were collected at the $L_\mathrm{2,3}$ edges of Ni.\@ 
To ensure measurements on freshly prepared and shiny surfaces, the samples were cleaved shortly before the experiment in a preparation chamber with a base pressure of $\approx$ 1.6 $\times$ 10$^{-9}$ mbar and then transferred in ultrahigh vacuum to the NEXAFS chamber with a pressure of 1$\times$10$^{-10}$ mbar.\@ The surface cleanliness of the samples before and during data collection was monitored with x-ray photoemission spectroscopy. Using linearly polarized light, the spectra were taken in total electron yield (TEY) mode. Photon energy calibration was ensured by adjusting the Ni $L_\mathrm{3}$ peak position measured on a NiO standard before and after each NEXAFS scan to the established peak position. While the in-plane spectra with the polarization $E$ parallel to the $a$ and $c$ direction ($E||a$ and $E||c$) were directly obtained using the corresponding normal incidence alignment, out-of-plane spectra ($E||b$) were determined by measuring in a grazing incidence setup (with a grazing angle of 60$^{\circ}$)\@ and subsequently extrapolating the grazing spectra to 90$^{\circ}$ with the aid of the relevant in-plane spectrum.\@ With regard to an optimal presentation, only the important Ni $L_3$ edges recorded at different temperatures in the TEY mode along the \textit{a}, \textit{b} and \textit{c} directions will be presented in the main text. For completeness a full NEXAFS spectrum measured over the whole energy range is shown as an example in Fig.\@ S8 in the SM \cite{SM}.

\section{Results and Discussion}

\subsection{XRD: Structure determination and refinement}

The structural properties of Ta$_\mathrm{2}$NiSe$_\mathrm{5}$, Ta$_\mathrm{2}$NiS$_\mathrm{5}$, and Ta$_\mathrm{2}$Ni$_\mathrm{0.93}$Co$_\mathrm{0.07}$Se$_\mathrm{5}$, have been studied in detail as a function of temperature.\@ For Ta$_\mathrm{2}$NiSe$_\mathrm{5}$, the high-temperature structure was refined in the orthorhombic SG \textit{Cmcm} as illustrated on the left side of Fig.\@ \ref{fig: hybridization model},\@ while the low-temperature phase was refined in the monoclinic SG \textit{C2/c}.\@ Ta$_\mathrm{2}$Ni$_\mathrm{0.93}$Co$_\mathrm{0.07}$Se$_\mathrm{5}$ and Ta$_\mathrm{2}$NiS$_\mathrm{5}$ share the same orthorhombic high-temperature structure as Ta$_\mathrm{2}$NiSe$_\mathrm{5}$ with slightly smaller lattice parameters---more structural details on the lattice parameters and atomic positions at the highest and lowest measured temperature are given in the SM \cite{SM}.

\begin{figure}[t]
\centering
\hspace*{-5mm}
\includegraphics[scale=0.2]{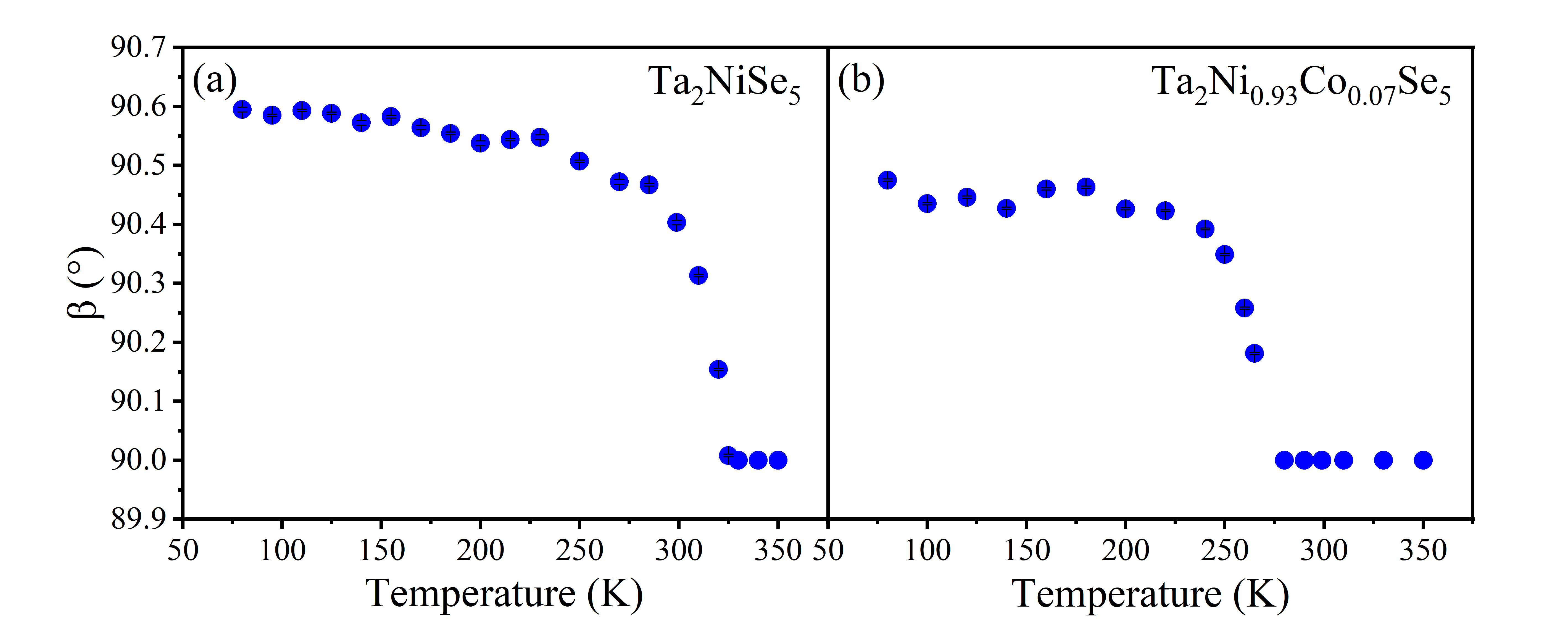}
\caption{Monoclinic order parameter $\beta$ of (a) Ta$_\mathrm{2}$NiSe$_\mathrm{5}$ and (b) Ta$_\mathrm{2}$Ni$_\mathrm{0.93}$Co$_\mathrm{0.07}$Se$_\mathrm{5}$  as a function of temperature. Errors shown are statistical errors from the refinements. Statistical error bars are smaller than the symbol sizes.}
\label{fig:Beta of Ta2NiSe5 and Ta2NiCoSe5}
\end{figure}

\begin{figure*}[t]
\centering
\hspace*{-4mm}
\includegraphics[scale=0.67]{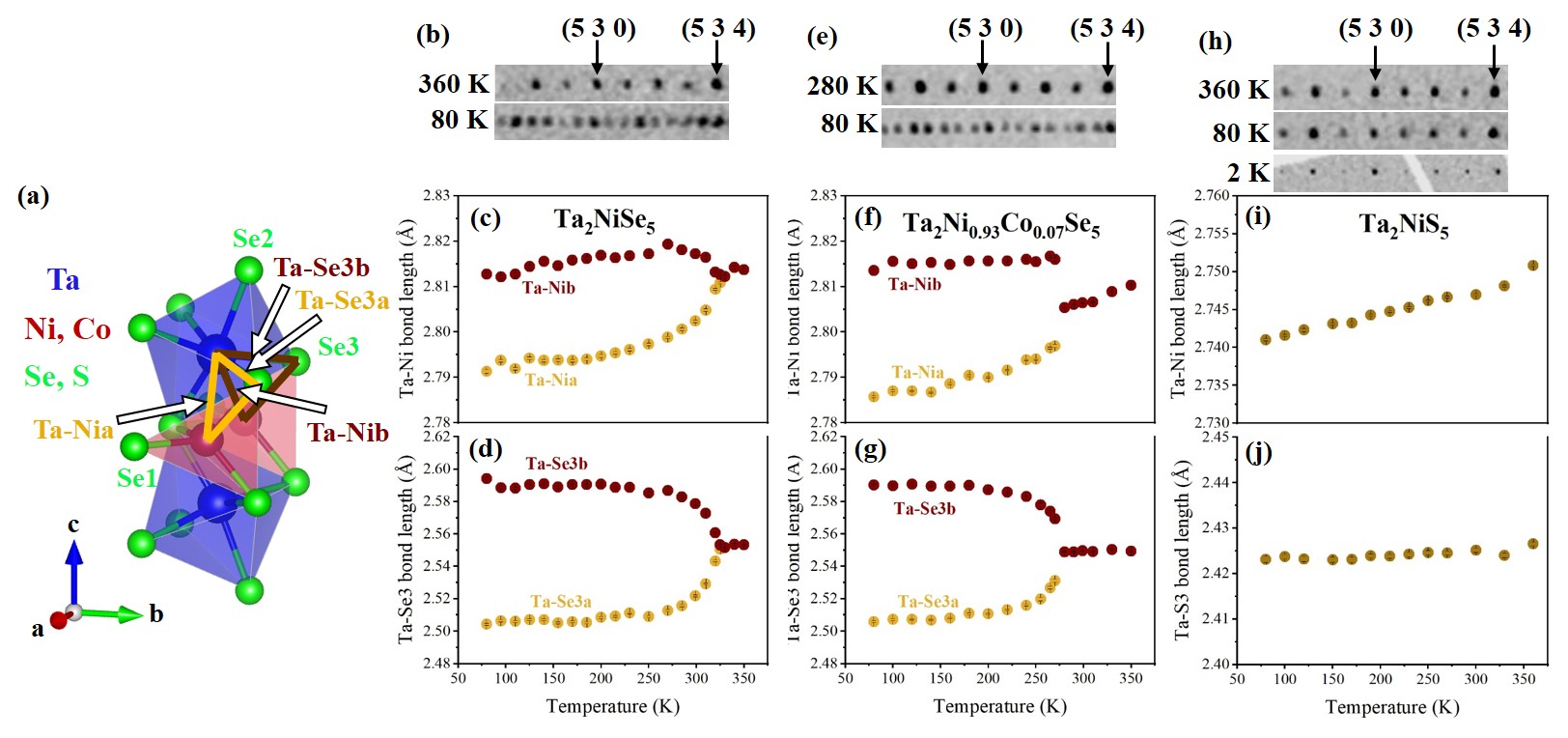}
\caption{(a) Part of the crystal structure in the monoclinic phase at 80 K. The yellow (brown) arrows represent the bonds that become shorter (longer) below the transition. Interatomic Ta-Ni and Ta-Se3/S3 distances in Ta$_\mathrm{2}$NiSe$_\mathrm{5}$ (c and d),
Ta$_\mathrm{2}$Ni$_\mathrm{0.93}$Co$_\mathrm{0.07}$Se$_\mathrm{5}$ (f and g) and Ta$_\mathrm{2}$NiS$_\mathrm{5}$ (i and j) as a function of temperature. Error bars reflect the statistical errors from the refinement and are smaller than the symbol sizes. A clear splitting of the Ta-Ni and Ta-Se3 bond lengths can be observed for Ta$_\mathrm{2}$NiSe$_\mathrm{5}$ and Ta$_\mathrm{2}$Ni$_\mathrm{0.93}$Co$_\mathrm{0.07}$Se$_\mathrm{5}$ which is totally absent for Ta$_\mathrm{2}$NiS$_\mathrm{5}$ where only a small reduction with decreasing temperature is found, simply reflecting the standard thermal expansion behavior. Panels (b) and (e) are parts of the x-ray precession images for the reciprocal (5 3 $l$) series ($l$ is changing from 4 to -3) reconstructed from single-crystal XRD data collected at 360 and 80 K for Ta$_\mathrm{2}$NiSe$_\mathrm{5}$, and at 280 and 80 K for Ta$_\mathrm{2}$Ni$_\mathrm{0.93}$Co$_\mathrm{0.07}$Se$_\mathrm{5}$ show the splitting of the reflections below the orthorhombic-to-monoclinic phase transition. Panel (h) shows the precession images for Ta$_\mathrm{2}$NiS$_\mathrm{5}$ again for the reciprocal (5 3 $l$) series ($l$ is from 4 to -3) at 360 and 80 K from our XRD data, and for the reciprocal (5 1 $l$) series ($l$ is from 4 to 11) at 2 K from Synchrotron XRD data, where no splitting of the reflections is seen in the whole temperature range. More details and complete precession plots are given in the SM.\@}

\label{fig: interatomic distances of all compounds}
\end{figure*}

Fig.\@ \ref{fig:Beta of Ta2NiSe5 and Ta2NiCoSe5} shows the $T$-dependent behavior of the monoclinic order parameter $\beta$ for (a) Ta$_\mathrm{2}$NiSe$_\mathrm{5}$  and (b) Ta$_\mathrm{2}$Ni$_\mathrm{0.93}$Co$_\mathrm{0.07}$Se$_\mathrm{5}$.\@ Upon cooling, a continuously increasing difference from 90$^{\circ}$ coupled to the second order structural phase transition is evident below $T_\mathrm{C}$ and finally $\beta$ reaches values of 90.59$^{\circ}$ for Ta$_\mathrm{2}$NiSe$_\mathrm{5}$ and 90.48$^{\circ}$ for Ta$_\mathrm{2}$Ni$_\mathrm{0.93}$Co$_\mathrm{0.07}$Se$_\mathrm{5}$  at 80 K.\@ The transition temperatures can be determined to $\approx$ 327 K for Ta$_\mathrm{2}$NiSe$_\mathrm{5}$ and $\approx$ 265 K for Ta$_\mathrm{2}$Ni$_\mathrm{0.93}$Co$_\mathrm{0.07}$Se$_\mathrm{5}$, which is in good agreement with $T_\mathrm{C}$ values reported in literature \cite{sunshine_24_1985, SALVO_116_1986, Mitsuoka_89_2020, Sano_30_2020}. As representatives, panels (b) and (e) in Fig. \ref{fig: interatomic distances of all compounds} illustrate a series of (5 3 $l$) reflections taken from precession images for the reciprocal ($h$ 3 $l$) plane reconstructed from single-crystal XRD data collected at 360 and 80 K for Ta$_\mathrm{2}$NiSe$_\mathrm{5}$ and at 280 and 80 K for Ta$_\mathrm{2}$Ni$_\mathrm{0.93}$Co$_\mathrm{0.07}$Se$_\mathrm{5}$ and demonstrate the splitting of the reflections at and below the orthorhombic-to-monoclinic phase transition.

For Ta$_\mathrm{2}$NiS$_\mathrm{5}$, across the entire measured temperature range from 360 to 80 K, the lattice parameters (and consequently the unit cell volume) change smoothly, showing a gradual decrease with decreasing temperature. However, throughout the entire reciprocal lattice, and particularly in the ($h$ $k$ $l$) layers with $k = 0,1,2,3,\dots$\@ no evidence of reflection splitting is observed.\@ This clearly indicates that the compound does not undergo any phase transition down to 80 K. To ensure that no phase transition occurs below 80 K, we conducted additional synchrotron XRD measurements down to 2 K.\@ None of the accessible reflections in the ($h$ $k$ $l$) planes with $k = 0,1,2,3, \dots$ showed any signs of a splitting or a broadening expected in the case of a twinned crystal below the phase transition. This is exemplified, for instance, by the series of (5 3 $l$) reflections in panel (h) of Fig.\@ \ref{fig: interatomic distances of all compounds} which were taken from the reciprocal ($h$ 3 $l$) plane reconstructed from single-crystal XRD data collected at 360 and 80 K,\@ as well as from synchrotron XRD data collected at 2 K for Ta$_\mathrm{2}$NiS$_\mathrm{5}$ (full precession figures are shown in Figs.\@ S3, S4, S5, and S6 in the SM \cite{SM}). Thus, we can rule out any structural phase transition in our samples down to 2 K. These findings are in line with the resistivity measurements of Ref.\@ \cite{Lu_8_2017}, which also reported no phase transition in Ta$_\mathrm{2}$NiS$_\mathrm{5}$ down to 2 K.\@ The conflicting reports of a structural instability \cite{Ye_110_2024} and even a structural phase transition \cite{Ye_104_2021} from Raman experiments, supported by first principle calculations \cite{Windgätter_7_2021}, may arise from the fact that XRD and Raman probe different sample volumes with varying coherence lengths. Furthermore, the Raman results might suggest a tendency towards a phase transition that is never {\it fully} realized and long-range ordered, and which, therefore, remains undetected in diffraction experiments.

\begin{figure}[b]
\centering
\includegraphics[scale=0.35]{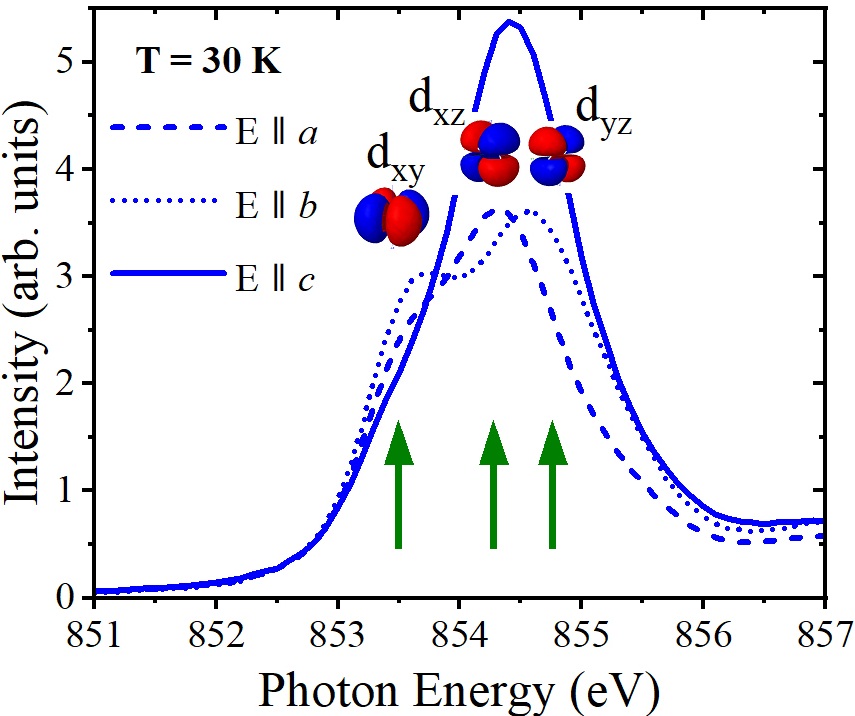}
\caption{Comparison of Ni $L_3$ NEXAFS spectra of Ta$_\mathrm{2}$NiSe$_\mathrm{5}$ recorded at 30 K along \textit{a}, \textit{b}, and  \textit{c}.\@ The in-plane $d_\mathrm{xy}$ orbitals are observed in the lower energy range, below 854 eV, with spectral contributions along $E{\parallel}a$ and $E{\parallel}b$.\@ In contrast, out-of-plane $d_\mathrm{xz}$ orbitals show contributions along $E{\parallel}a$ and $E{\parallel}c$,\@ whereas the $d_\mathrm{yz}$ orbitals exhibit contributions along $E{\parallel}b$ and $E{\parallel}c$,\@ and are found at slightly higher energies.}
\label{fig:NEXAFS at 30 K along three directions for Ta2NiSe5}
\end{figure}

With the transition from the orthorhombic (\textit{Cmcm}) to the monoclinic (\textit{C2/c}) phase observed in Ta$_\mathrm{2}$NiSe$_\mathrm{5}$ and Ta$_\mathrm{2}$Ni$_\mathrm{0.93}$Co$_\mathrm{0.07}$Se$_\mathrm{5}$,\@ the two mirror symmetries shown in Fig.\@ \ref{fig: hybridization model} (left) are broken, resulting in the splitting of the Ni and Se atomic positions (see Fig.\@ \ref{fig: interatomic distances of all compounds} and Table S2 in the SM \cite{SM}): Above $T_\mathrm{C}$, the Ta-Ni bond length remains almost unchanged. However, below $T_\mathrm{C}$, this bond splits, and one bond, Ta-Nib, increases significantly, while the other, Ta-Nia, decreases markedly [Fig. \ref{fig: interatomic distances of all compounds} (c) and (f)].\@ A similar pattern is observed for the splitting of the Ta-Se3 distances, as shown in Fig.\@ \ref{fig: interatomic distances of all compounds} (d) and (g).\@ From this we conclude that the breaking of mirror symmetries and the subsequent shortening of the Ta-Nia and Ta-Se3a bond lengths substantially enhance the hybridization between the relevant Ta, Nia, and Se3a orbital states. Conversely, the lengthening of the Ta-Nib and Ta-Se3b bonds weakens the hybridization between the Ta,\@ Nib,\@ and Se3b orbital states.\@ This interplay of decreased and increased bond lengths, together with the observed changes in the $\beta$ angle, aligns with the shearing of the system already reported in the literature \cite{Watson_2_2020}. It is worth mentioning that a minor lattice contraction can be seen with Co substitution at the Ni site (see Table S1 in the SM \cite{SM}),\@ due to the fact that Co is the immediate neighbor of Ni in the periodic table and therefore has a slightly smaller atomic radius and one less electron in the $3d$ shell. Due to the minimal magnitude of this effect, along with the negligible changes in the relevant Ta-Ni/Co and Ni/Co-Se bond distances, a change in the valence or spin state can be ruled out. Furthermore, it can be seen from Figs.\@ \ref{fig: interatomic distances of all compounds} (c), (d), (f), and (g),\@ that the bond distances split, but the average distance---which determines the volume around the Ni atom---remains the same above and below the transition. This strongly suggests a change in hybridization rather than a change in the valence or spin state.

\begin{figure*}[t]
\includegraphics[scale=0.29]{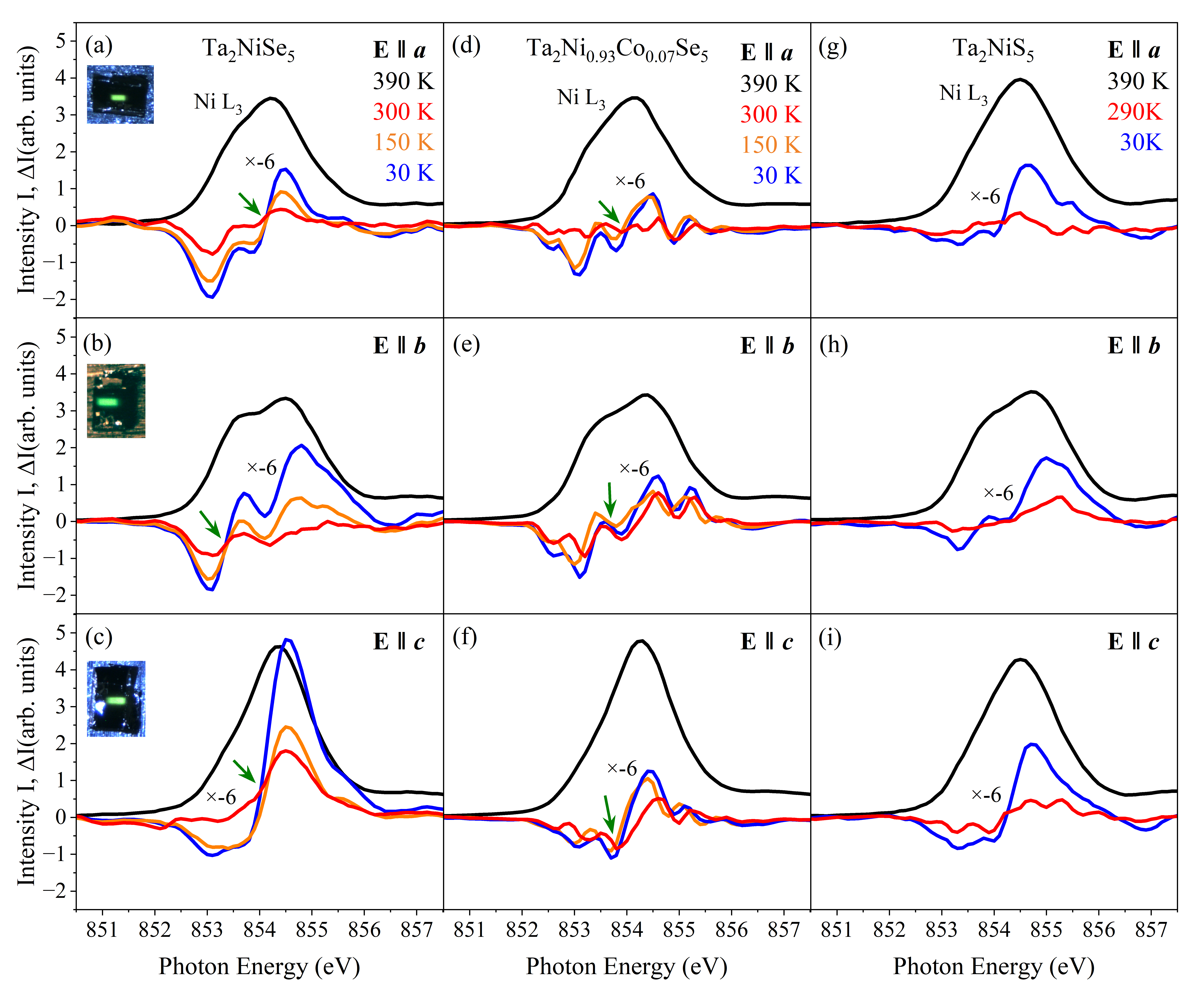}
\caption{NEXAFS spectra of Ta$_\mathrm{2}$NiSe$_\mathrm{5}$ (left), Ta$_\mathrm{2}$Ni$_\mathrm{0.93}$Co$_\mathrm{0.07}$Se$_\mathrm{5}$ (middle) and Ta$_\mathrm{2}$NiS$_\mathrm{5}$ (right) with the beam parallel to \textit{a}, \textit{b}, and \textit{c} directions. Shown is the spectrum taken at 390 K and the difference $\Delta$I = I(390 K) - I (T) between the spectra taken at 390 K and at the respective temperatures given in the graph. Note that $\Delta$I is multiplied by a factor of -6 and that a negative area means loss of spectral weight relative to 390 K. The green arrows point to the isosbestic point. The insets in the upper left corner of figures (a), (b), and (c) illustrate the polarization of light with respect to the sample’s orientation. A significant charge transfer from orbitals with exclusively in-plane character, i.e., d$_\mathrm{xy}$, to  orbitals with in- and out-of-plane character, i.e., d$_\mathrm{yz}$, is found below T$_C$ for Ta$_\mathrm{2}$NiSe$_\mathrm{5}$ and Ta$_\mathrm{2}$Ni$_\mathrm{0.93}$Co$_\mathrm{0.07}$Se$_\mathrm{5}$, while only a minor charge transfer of this type is found between in-plane and out-of-plane orbitals for Ta$_\mathrm{2}$NiS$_\mathrm{5}$. For Ta$_\mathrm{2}$NiS$_\mathrm{5}$, temperature effects seem to play the dominant role.}
\label{fig:all NEXAFS spectra}
\end{figure*}

In Ta$_\mathrm{2}$NiS$_\mathrm{5}$, on the other hand, the $T$ dependence of the Ta-Ni and Ta-S3 bond lengths displays a small reduction with decreasing temperature, thereby, simply reflecting the standard thermal expansion behavior. In the whole temperature range, however, no changes in the bond distances signaling a symmetry breaking are found [see Fig.\@ \ref{fig: interatomic distances of all compounds} (i) and (j)].\@

\subsection{NEXAFS: Electronic properties of Ta\texorpdfstring{$_{2}$}{2}NiSe\texorpdfstring{$_{5}$}{5}, Ta\texorpdfstring{$_{2}$}{2}Ni\texorpdfstring{$_{0.93}$}{0.93}Co\texorpdfstring{$_{0.07}$}{0.07}Se\texorpdfstring{$_{5}$}{5}, and Ta\texorpdfstring{$_{2}$}{2}NiS\texorpdfstring{$_{5}$}{5} }

To investigate the relationship between the observed electronic and structural effects, particularly in relation to the orbital physics at the electronically dominant Ni sites, NEXAFS experiments were conducted in TEY mode at the Ni $L_{3}$ edge. In all the compounds studied here, the $e$-type orbitals ($d_\mathrm{x^2-y^2}$ and $d_\mathrm{3z^2-r^2}$) located at lower energies are fully occupied and, therefore, spectroscopically silent. In contrast, the $t_\mathrm{2}$ orbitals ($d_\mathrm{xy,xz,yz}$) are active at the Fermi level and play a crucial role in the material's electronic behavior \cite{Merz_85_2016}.

Fig.\@ \ref{fig:NEXAFS at 30 K along three directions for Ta2NiSe5} shows Ta$_\mathrm{2}$NiSe$_\mathrm{5}$ NEXAFS spectra recorded in the monoclinic phase at 30 K along the crystallographic \textit{a}, \textit{b}, and \textit{c} directions at the Ni $L_\mathrm{3}$ edge. The distinct behavior of the three spectra enables the identification of the orbitals involved and clearly demonstrates that the in-plane $d_\mathrm{xy}$ orbitals are dominating in the lower energy range below 854 eV with their spectral contributions along $E{\parallel}a$ and $E{\parallel}b$, while the orbitals with out-of-plane character are primarily probed at slightly higher energies with spectral contributions along $E{\parallel}a$ and $E{\parallel}c$ for $d_\mathrm{xz}$ and along $E{\parallel}b$ and $E{\parallel}c$ for $d_\mathrm{yz}$ orbitals.

\begin{figure}[b]
\centering
\includegraphics[scale=0.7]{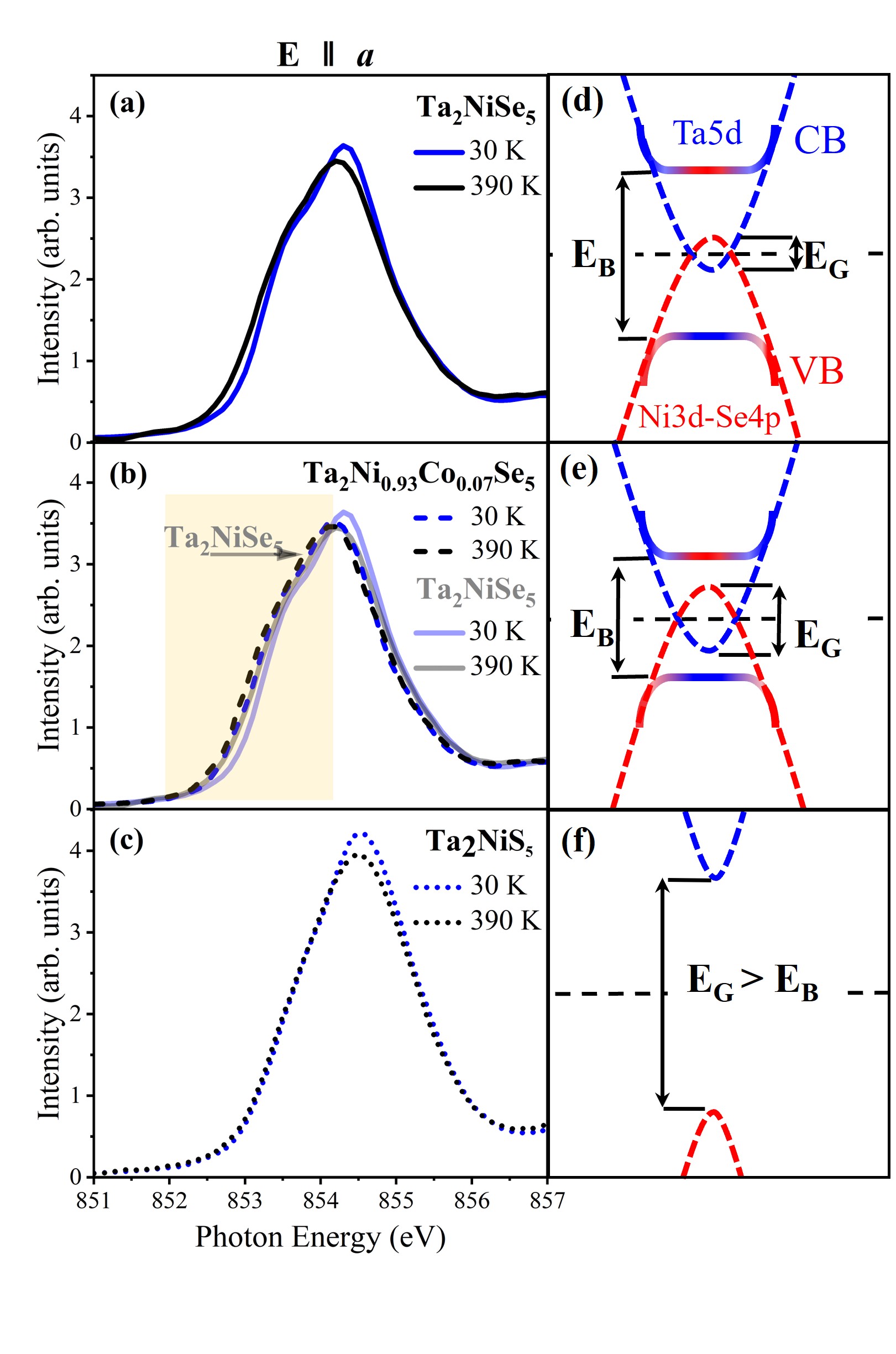}
\caption{Comparison of Ni L$_3$ NEXAFS spectra of (a) Ta$_\mathrm{2}$NiSe$_\mathrm{5}$, (b) Ta$_\mathrm{2}$Ni$_\mathrm{0.93}$Co$_\mathrm{0.07}$Se$_\mathrm{5}$, and (c) Ta$_\mathrm{2}$NiS$_\mathrm{5}$ recorded at 390 and 30 K along the \textit{a} direction. In the plot for Ta$_\mathrm{2}$Ni$_\mathrm{0.93}$Co$_\mathrm{0.07}$Se$_\mathrm{5}$, the spectra of Ta$_\mathrm{2}$NiSe$_\mathrm{5}$ are additionally depicted in gray. The right panel of the figure represents sketches for the band structure of the three compounds before and after the formation of the excitonic insulator state as derived from our NEXAFS data.}

\label{fig:NEXAFS spectra of the 3 compounds at 30 and 390K}
\end{figure}

In order to emphasize the orbital occupation redistribution across the investigated temperature range, we show in Fig.\@ \ref{fig:all NEXAFS spectra} the high-temperature NEXAFS spectra measured along \textit{a}, \textit{b}, and \textit{c} directions at the Ni $L_\mathrm{3}$ edge for Ta$_\mathrm{2}$NiSe$_\mathrm{5}$ (left), Ta$_\mathrm{2}$Ni$_\mathrm{0.93}$Co$_\mathrm{0.07}$Se$_\mathrm{5}$ (middle), and Ta$_\mathrm{2}$NiS$_\mathrm{5}$ (right) together with the corresponding temperature-dependent  difference spectra $\Delta$I = I(390 K) - I (T).\@ The insets in the upper left corners of panels (a), (b), and (c) illustrate the orientation of the beam relative to the sample. For Ta$_\mathrm{2}$NiSe$_\mathrm{5}$,\@ spectra were recorded in the orthorhombic phase at 390 K and in the monoclinic phase at 290, 150, and 30 K. In this compound, the spectra exhibit strong anisotropy with respect to the crystallographic directions. Additionally, an isosbestic point (ISP) where all temperature-dependent spectra coincide and which is indicated by an arrow, is observed around 854 eV for all three directions, with a notable temperature-dependent transfer of spectral weight from the low-energy to the high-energy side of the ISP as the temperature decreases below T$_C$.\@  This transfer of spectral weight signals that hole carriers are being shifted away from $E_F$ (i.e., from the region below the ISP) towards the first unoccupied states above $E_F$ (i.e., to the region above the ISP).\@

Comparing the directional dependence in Figs.\@ \ref{fig:all NEXAFS spectra} (a), (b), and (c), it is evident that the holes are removed from the in-plane orbitals at $E_F$,\@ which are predominantly probed below the ISP for $E{\parallel}a$ and $E{\parallel}b$  [Fig.\@ \ref{fig:all NEXAFS spectra} (a) and (b)].\@ These holes are transferred to orbitals with out-of-plane character, which are primarily probed above the ISP for $E{\parallel}b$ and $E{\parallel}c$ [Fig.\@ \ref{fig:all NEXAFS spectra} (b) and(c)],\@ as illustrated in Fig.\@ \ref{fig:NEXAFS at 30 K along three directions for Ta2NiSe5}.\@ In effect, this indicates a substantial charge transfer from in-plane $d_\mathrm{xy}$ to out-of-plane $d_\mathrm{yz}$ orbitals below the transition temperature of Ta$_\mathrm{2}$NiSe$_\mathrm{5}$.\@ It should be noted that the small change in the order parameter $\beta$,\@ as shown in Fig.\@ \ref{fig:Beta of Ta2NiSe5 and Ta2NiCoSe5},\@ which is associated with the mirror symmetry breaking, is solely in-plane and too minor to be detected in NEXAFS experiments. However, what is clearly observed is a transfer of spectral weight between different directions, which is only possible if the orbital character changes from in-plane to out-of-plane as previously described (see also discussion of Fig.\@ S9 in the SM \cite{SM}).\@ Furthermore, the spectral shape indicates a Ni$^{3+}$ state, and---consistent with our structural data---a temperature-dependent change in the valence or spin state can be ruled out due to the minimal modifications observed in the multiplet structures (see discussion in the SM \cite{SM} and Refs.\@ \cite{Merz_33_2021, Kleiner_7_2015, Kleiner_168_2021}).

Ni L$_\mathrm{3}$ spectra of Ta$_\mathrm{2}$Ni$_\mathrm{0.93}$Co$_\mathrm{0.07}$Se$_\mathrm{5}$, measured at 390 and 290 K (above the transition) and at 150 K and 30 K (below the transition), exhibit a similar but less pronounced behavior compared to Ta$_\mathrm{2}$NiSe$_\mathrm{5}$, as can be seen in Figs.\@ \ref{fig:all NEXAFS spectra} and \ref{fig:NEXAFS spectra of the 3 compounds at 30 and 390K}.\@ With Co doping, the material becomes a more distinct semimetal, characterized by a larger overlap between the conduction and valence bands, as depicted in the band structure sketch in panel (e) of Fig.\@ \ref{fig:NEXAFS spectra of the 3 compounds at 30 and 390K}.\@ As a result, the Ni L$_\mathrm{3}$ spectra of the  high-$T$ phase of Ta$_\mathrm{2}$Ni$_\mathrm{0.93}$Co$_\mathrm{0.07}$Se$_\mathrm{5}$ show in the low-energy region below the ISP a higher spectral weight compared to Ta$_\mathrm{2}$NiSe$_\mathrm{5}$ (see the yellow shaded area in panel (b) of Fig.\@ \ref{fig:NEXAFS spectra of the 3 compounds at 30 and 390K}).\@ Below the transition, both compounds behave as EIs.\@ Yet, Ta$_\mathrm{2}$Ni$_\mathrm{0.93}$Co$_\mathrm{0.07}$Se$_\mathrm{5}$ has a larger negative energy gap, leading to a stronger screening and, thus, to a reduced transition temperature and a smaller number of excitons, and as a consequence results in less spectral weight transfer. Therefore, fewer hole states move from the low-energy to the high-energy side of the ISP at the onset of the EI.\@ A sketch of the corresponding band structures derived from our NEXAFS data is illustrated for all three systems in (d), (e), and (f) of
 Fig.\@ \ref{fig:NEXAFS spectra of the 3 compounds at 30 and 390K}.

For Ta$_\mathrm{2}$NiSe$_\mathrm{5}$, the shift between the 390 and 30 K spectra can be determined from Fig.\@ \ref{fig:NEXAFS spectra of the 3 compounds at 30 and 390K} and amounts to $\approx$ 110 meV.\@ This value is roughly consistent with the optical excitation gap E$_{op}$ $\sim$ 0.16 eV below T$_C$ and the estimated exciton binding energy E$_B$ $\sim$ 0.17 eV from a previous ARPES study \cite{Lu_8_2017}.\@ In the case of Ta$_\mathrm{2}$Ni$_\mathrm{0.93}$Co$_\mathrm{0.07}$Se$_\mathrm{5}$, the shift and with it its exciton binding energy  is strongly reduced and can be estimated to be $\approx$ 50 meV.\@ Notably, the respective spectra above the transition (at 290 K and 390 K) are nearly on top of each other, as do the spectra below the transition (at 150 K and 30 K) and we can conclude that the observed gap opens at the transition to the EI state.\@ The slight difference between the spectra above (and below) $T_C$ can be attributed to phononic broadening, which increases at higher temperatures: In other words, as the temperature rises, the spectral width slightly broadens while its maximum intensity decreases.\@ This trend is consistent across all individual crystallographic directions and is observed in the spectra of Ta$_\mathrm{2}$NiSe$_\mathrm{5}$ and Ta$_\mathrm{2}$NiS$_\mathrm{5}$ (see below) as well.

\begin{figure*}
\centering
\hspace*{-5mm}
\includegraphics[scale=0.57]{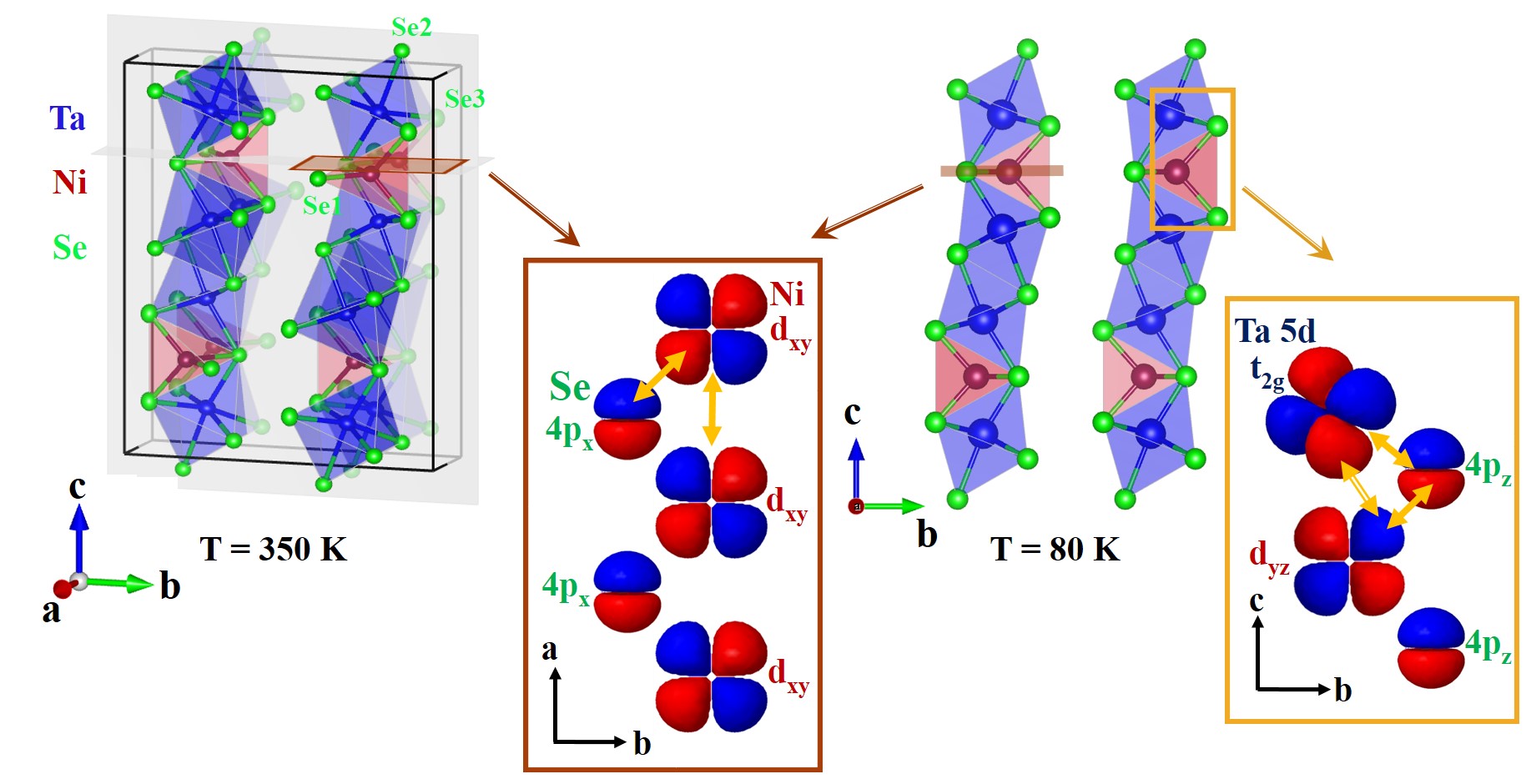}
\caption{Orbital model for Ta$_\mathrm{2}$NiSe$_\mathrm{5}$: (a) In the orthorhombic phase at 350 K, and (b) in the monoclinic phase at 80 K. In both sketches the Ni $d_{xy}$ and $d_{yz}$ orbitals symbolize the states relevant for the hybridization in the NiSe$_4$ tetrahdron. Similarly, the Ta $5d_{1/ \sqrt 2 \times (d_{xz} + d_{yz})}$ orbitals (coordinates were defined with respect to the unit cell) pointing between the Se neighbors and towards the Ni atoms symbolize the states relevant for the hybridization in the TaSe$_6$ octahedron and for the direct Ta-Ni bond. Among the Ta $5d$ $t_2$ states, these orbitals are the only ones that can effectively hybridize simultaneously with both the Ni $4d_{yz}$ and Se $4p_z$ orbitals. In the orthorhombic phase, hybridization between Ni and Ta states are forbidden due to their opposite mirror parity, while in the monoclinic phase, two mirror symmetries are broken, permitting hybridization between the relevant Ta, Ni, and Se orbitals. Note, the perspective view and the projections along the main directions change between (a) and (b),\@ as indicated by the red and orange rectangles in the figures. } 
\label{fig: hybridization model}
\end{figure*}

With S substitution,\@ the semiconducting band gap widens, as shown in the sketch (f) of Fig. \ref{fig:NEXAFS spectra of the 3 compounds at 30 and 390K}. The Ni L$_\mathrm{3}$ spectra of Ta$_\mathrm{2}$NiS$_\mathrm{5}$, measured at 390 K, 290 K, and 30 K along the three crystallographic directions are almost isotropic (see panel (c) of Fig.\@ \ref{fig:NEXAFS spectra of the 3 compounds at 30 and 390K}) and from the directional-dependent spectra in Fig. \ref{fig:all NEXAFS spectra} (g), (h), and (i), a much smaller spectral weight transfer between in-plane and out-of-plane spectra is observed: In contrast to the other two compounds where a clear charge transfer between $d_\mathrm{xy}$ and $d_\mathrm{yz}$ orbitals is found, the spectral weight is for Ta$_\mathrm{2}$NiS$_\mathrm{5}$ predominantly transferred from the low-energy to the high-energy region within the spectra of the same directional character, i.~e., from the low-energy to the high-energy side of the respective $E{\parallel}a$, $E{\parallel}b$, and $E{\parallel}c$ spectra but for example not {\it between} $E{\parallel}a$ and $E{\parallel}b$ or $E{\parallel}c$.\@ This indicates that the observed kind of spectral weight transfer likely originates from temperature effects rather than from any intrinsic EI mechanism. Furthermore, no shift is observed between the 390 and 30 K $E{\parallel}a$ spectra in Fig.\@ \ref{fig:NEXAFS spectra of the 3 compounds at 30 and 390K} (c), providing no evidence for an EI binding energy and strongly suggests the absence of a stable long-range EI phase in Ta$_\mathrm{2}$NiS$_\mathrm{5}$.\@ However, a small, potentially fluctuating EI component cannot be entirely ruled out based on our NEXAFS data, and might even be consistent with the EI instability recently reported in Raman experiments \cite{Ye_110_2024, Ye_104_2021}.\@

In the following, we will combine our structural and electronic findings to construct a comprehensive orbital picture: In the orthorhombic phase above T$_C$,\@ Ni and Ta states exhibit different mirror symmetries, as discussed above and in \cite{Watson_2_2020}, which prevents any hybridization between them. Our NEXAFS measurements at the Ni edge indicate that in the high-temperature phase $d_{xy}$ orbitals dominate at $E_F$.\@ Consequently, hybridization is mainly possible along the crystallographic \textit{a} direction, specifically between the $3d_{xy}$ orbitals of Ni and the $4p_x$ orbitals of Se atoms, as illustrated in the left panel of Fig.\@ \ref{fig: hybridization model}.\@ In contrast, below the transition to the monoclinic phase, the mirror symmetries both parallel and perpendicular to the chain direction are broken as exemplified in our structural data. This symmetry breaking enables hybridization between the relevant Ta, Ni, and Se orbitals in the low-temperature phase and the shortening of the Ta-Nia and Ta-Se3a bond lengths (see Fig.\@ \ref{fig: interatomic distances of all compounds}) clearly manifests the significantly enhanced hybridization. 
Our NEXAFS data, along with the structural findings, reveal that upon cooling, holes move from in-plane orbitals ($d_{xy}$) to orbitals with out-of-plane components ($d_{xz}$ and $d_{yz}$), with the $d_{yz}$ orbitals becoming the dominant states just above E$_{F}$.\@ The right panel of Fig.\@ \ref{fig: hybridization model} summarizes this hybridization scenario where Ni $d_{yz}$ orbitals hybridize \textit{directly} with lower-energy Ta $t_{2g}$ $1/ \sqrt 2 \times (d_{xz} + d_{yz})$ orbitals as well as \textit{indirectly} mediated through hybridization with $p_z$ orbitals of Se atoms (as explained in the caption of Fig.\@ \ref{fig: hybridization model}),\@ and it seems to be precisely this hybridization among these three elements that stabilizes the EI state in this system. According to our XRD and NEXAFS results, this trend is most pronounced for Ta$_\mathrm{2}$NiSe$_\mathrm{5}$ and significantly reduced for Ta$_\mathrm{2}$Ni$_\mathrm{0.93}$Co$_\mathrm{0.07}$Se$_\mathrm{5}$ and almost absent in the case of Ta$_\mathrm{2}$NiS$_\mathrm{5}$.

This is a key feature of the EI, where we do not observe a simple or compensated shift of the valence and conduction bands relative to the Fermi level to open a band gap, as one might expect in conventional scenarios.\@ Nor is the gap driven by strong correlation effects of charge carriers of the identical type residing in the same orbital states, as described by the Hubbard model. Instead, the band gap emerges due to symmetry breaking and the hybridization of Ta, Ni, and Se states that are forbidden by the symmetry of the high-temperature phase.

Our data not only confirm the occurrence of symmetry breaking and orbital state hybridization, but also allow us to identify the orbitals involved in this process. Furthermore, our results are completely in line with the expected dome-shaped trend characteristic of an EI:\@ The formation of the EI is most pronounced in Ta$_\mathrm{2}$NiSe$_\mathrm{5}$,\@ but as Co is substituted, shifting the material towards the semi-metallic side, the emergence of the EI is rapidly suppressed due to the stronger screening of the excitons (Ta$_\mathrm{2}$Ni$_\mathrm{0.93}$Co$_\mathrm{0.07}$Se$_\mathrm{5}$).\@ On the contrary, with increasing S content the semiconducting band gap widens and the EI state is strongly suppressed. However, a certain resuming tendency to an EI state, likely driven by short-range fluctuations, cannot be entirely ruled out. Nonetheless, there is no evidence in our data for symmetry breaking, for an EI binding energy, or changes in the hybridization properties in Ta$_\mathrm{2}$NiS$_\mathrm{5}$.\@

While our experiments alone, similar to all previous investigations, do not offer definitive smoking-gun proof for an EI state in Ta$_\mathrm{2}$NiSe$_\mathrm{5}$,\@ the likelihood of such a state is very high, especially when our findings are considered alongside the amount of results gathered from numerous other experiments. Yet, one of the most fundamental issues for the EI in Ta$_\mathrm{2}$NiSe$_\mathrm{5}$,\@ but in general as well, 
is that symmetry breaking is absolutely essential for allowing the electronic hybridization effect. At the same time, however, this unavoidably also leads to a structural distortion that is intrinsically coupled to the electronic one, resulting in a nearly inseparable chicken-and-egg problem, where it is difficult to ultimately disentangle whether the electronic or structural degrees of freedom are the driving force behind the phase transition. However, the data presented here unambiguously show that this is not merely a simple symmetry breaking, but that the low-$T$ phase is also accompanied by a change in the orbital symmetries of the Ta, Ni, and Se states at $E_F$.\@ 

This finding is crucial: it suggests that, in addition to enabling the hybridization, the structural transition is accompanied by significant changes in the orbital character, which play a vital role in the formation of the EI. This points to a strong, and potentially dominant, electronic component that is essential for the phase transition to an EI.\@

\section{Acknowledgments}
We gratefully acknowledge the KIT Light Source KARA, Karslruhe and the Karlsruhe Nano Micro Facility (KNMFi) for the provision of beamtime as well as the European Synchrotron Radiation Facility (ESRF) for provision of synchrotron radiation facilities. We are indepted to Siegmar Roth and Andre Beck at the Institute for Quantum Materials and Technologies, Karlsruhe Institute of Technology for their great technical assistance.

%\newpage

%\clearpage
%\bibliographystyle{apsrev4-2}
\bibliography{TaNiSe}% Produces the bibliography via BibTeX.
%\printbibliography

\section{Author Contributions}

M.M. conceived and managed the project. A.-A.H. and S.P. grew the single crystals. A.-A.H. performed EDX experiments. N.M., B.W., F.A., G.G., M.Y., M.L.T., and M.M. performed XRD experiments. N.M. and M.M. refined the XRD data and interpreted the results. P.N., F.G., A.G., S.S., N.M., and M.M. performed NEXAFS measurements.\@ P.N., S.S., and M.M. evaluated the NEXAFS spectra and M.M. interpreted the data. N.M. and M.M. wrote the manuscript with inputs from all authors.

\section{DATA AND MATERIALS AVAILABILITY}

All data is available in the manuscript or the supplementary information. Data taken at the ESFR, ID15B beamline are available under https://doi.esrf.fr/10.15151/ESRF-DC-2227830403.

\section{Competing interests}

The authors declare no competing interests.

\section{ADDITIONAL INFORMATION}

Supplementary Information accompanies this paper.

\end{document}